\documentclass[structabstract]{aa}
\usepackage{graphicx}
\usepackage{txfonts}
\input{epsf}
\usepackage{rotating}

\newcommand{\fbrg}{$F/F_{Br\gamma}$$^{\mathrm{a}}$}  
\newcommand{\mic}{\,$\mu$m}
\newcommand{\kms}{\,km s$^{-1}$}
\newcommand{\dg}{$^{\circ}$}
\newcommand{\wma}{\,W~m$^{-2}$~arcsec$^{-2}$}
\newcommand{\wmm}{\,W~m$^{-2}$~$\mu$m$^{-1}$}
\newcommand{\wm}{\,W~m$^{-2}$}
\newcommand{\Msolar}{\,M$_{\odot}$}
\newcommand{\Lsolar}{\,L$_{\odot}$}
\newcommand{\htwo}{H$_2$}
\newcommand{\feii}{[Fe~{\sc II}]}
\newcommand{\brg}{Br~$\gamma$}
\newcommand{\hi}{H~{\sc I}}
\newcommand{\hei}{He~{\sc I}}
\newcommand{\nai}{Na~{\sc I}}
\newcommand{\cai}{Ca~{\sc I}}
\newcommand{\mgi}{Mg~{\sc I}}
\newcommand{\ali}{Al~{\sc I}}
\newcommand{\tii}{Ti~{\sc I}}
\newcommand{\oi}{O~{\sc I}}
\newcommand{\sii}{S~{\sc II}}

\begin{document}

\title{A General Catalogue of Molecular Hydrogen Emission-Line Objects 
                       (MHOs) in Outflows from Young Stars          
        \thanks{http://www.jach.hawaii.edu/UKIRT/MHCat/}
       }

\authorrunning{Davis et al.}
\titlerunning{MHO Catalogue}

   \author{Christopher J. Davis\inst{1},
           Ryan Gell\inst{1,2},
           Tigran Khanzadyan\inst{3},
           Michael D. Smith\inst{4}, 
           Tim Jenness\inst{1}
          }

   \institute{
           Joint Astronomy Centre, 660 North A'oh\={o}k\={u} Place,
           University Park, Hilo, Hawaii 96720, USA \\
              \email{c.davis@jach.hawaii.edu}
       \and
           Faculty of Engineering, University of Victoria,
           Victoria BC, V8W 3P6, Canada 
       \and
           Centre for Astronomy, Department of Experimental Physics, National
           University of Ireland, Galway, Ireland
       \and
           Centre for Astrophysics \& Planetary Science,
           School of Physical Sciences, University of Kent,
           Canterbury CT2 7NR, U.K.
              }

 \date{Received July 1, 2008; accepted July 1, 2008}

  \abstract{
We present a catalogue of Molecular Hydrogen emission-line Objects
(MHOs) in outflows from young stars, most of which are deeply embedded.
All objects are identified in the near-infrared lines of molecular
hydrogen, all reside in the Milky Way, and all are associated with
jets or molecular outflows.  Objects in both low and
high-mass star forming regions are included.  This catalogue
complements the existing database of Herbig-Haro objects; indeed, for
completeness, HH objects that are detected in H$_2$ emission are
included in the MHO catalogue.

            }
   \maketitle


\section{Introduction}

For over 30 years, astronomers have been observing Herbig-Haro (HH)
objects, jets and outflows in star forming regions in the
near-infrared. The molecular hydrogen v=1-0~S(1) line at
2.122~\mic\ is a particularly powerful tracer of shock-excited
features in molecular outflows (e.g. Wilking et al. 1990; Garden,
Russell \& Burton 1990; Zealey et al. 1992; Gredel 1994; Davis \&
Eisl\"offel 1995; Zinnecker, McCaughrean \& Rayner 1998; Reipurth et
al. 1999; Eisl\"offel 2000; Stanke, McCaughrean \& Zinnecker 2002;
Caratti o Garatti et al. 2006; Walawender, Reipurth \& Bally 2009;
Davis et al. 2009). Although excited in a similar way to HH objects,
these molecular hydrogen emission-line features are often too deeply
embedded to be seen at optical wavelengths. They are thus not
classified as HH objects, which are strictly defined by optical
criteria, and are instead labelled in a rather hap-hazard way, often
with the authors' initials. In large on-line databases this can lead
to some ambiguity.

Our goal with this catalogue was therefore to develop a
self-consistent list of H$_2$ emission-line objects, in a
manner similar to that used so successfully for HH objects. With
guidance from the International Astronomical Union (IAU) Working Group
on Designations, we have adopted a scheme that simply lists objects
sequentially, although objects are grouped by region (see below). The
simple acronym ``MHO'', for Molecular Hydrogen emission-line Object,
is used to refer to these objects.  This acronym has been approved by
the IAU registry, and has been entered into the on-line Reference
Dictionary of Nomenclature of Celestial
Objects\footnote{http://cdsweb.u-strasbg.fr/cgi-bin/Dic?MHO}.

\begin{figure*}
\centering
  \includegraphics[width=14cm]{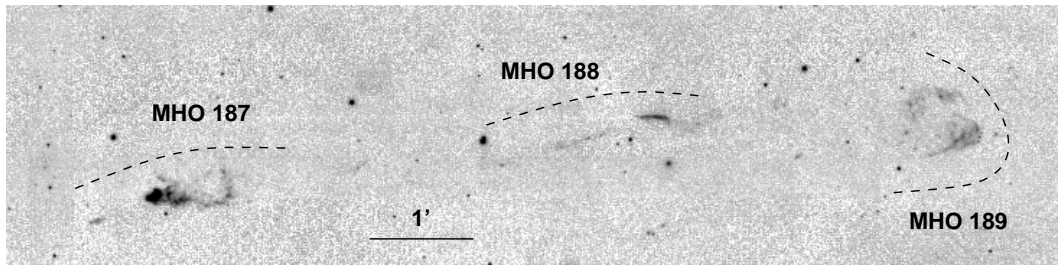}
\caption[]{\htwo\ (+ continuum) image of the MHO 187-189 outflow 
in Orion A.  Data from Davis et al. (2009).}
 \label{ori}
\end{figure*}

\section{What constitutes an MHO}

Only objects associated with outflows from Young Stellar Objects
(YSOs) and protostars are included in this catalogue. We do not list
outflows from evolved stars (AGB stars or Proto-Planetary Nebulae) or
extra-galactic sources. Also, objects should be spatially resolved;
unresolved emission-line regions associated with an accretion disk or
the base of an outflow (that were observed spectroscopically) are not
listed.

Since large-scale imaging surveys are now revealing tens or even
hundreds of objects in some regions (e.g. Stanke et al. 2002;
Khanzadyan et al. 2004b; Walawender et al. 2009; Davis et al. 2009),
spectroscopic confirmation of every feature is not usually practical
(although multi-object spectrographs that operate in the infrared will
certainly help in this regard). Therefore, to properly identify an MHO,
narrow-band molecular hydrogen images should be accompanied with
either adjacent narrow-band continuum images or (flux-scaled)
broad-band K images.  It is obviously important that these
shock-excited features be distinguished from wisps and knots of
continuum emission. Morphology alone should not be used to identify
MHOs, although the shape of an object may help distinguish features in
outflows from fluorescently excited emission regions, especially in
high-mass star-forming regions. If available, MHOs should have a
near-infrared spectrum consistent with thermal (shock) excitation,
rather than non-thermal (fluorescent) excitation (e.g. Gredel 1994;
Lorenzetti et al. 2002; Caratti o Garatti et al. 2006; Gianninni et
al. 2008).  Kinematic studies - either proper motion studies (Hodapp
1999; Davis et al. 2009) or high spectral-resolution line studies
(Carr 1993; Schwartz \& Greene 2003; Davis et al. 2004; Li et
al. 2008) - are also useful for distinguishing MHOs from what are
essentially stationary emission-line features in Photon-Dominated
Regions (PDRs).  The association of an MHO with a bipolar molecular
outflow, traced in (sub)millimeter molecular lines such as CO,
likewise confirms the dynamical nature and shock-excitation of
the object, and its association with a protostar (e.g. Yu et al 1999,
2000; Shepherd, Testi \& Stark 2003; Beuther, Schilke \& Stanke 2003;
Reipurth et al. 2004).

The MHOs listed in this catalogue have all been identified in the
near-infrared (1-2.5 \mic ) lines of molecular hydrogen. Objects
detected only in other near-IR lines (e.g. [FeII]) are not included.
We also exclude objects observed only in the UV or mid-infrared
(e.g. with the {\em Spitzer Space Telescope}).  If an object is
subsequently detected in molecular hydrogen line emission in the
near-IR, it will be included in the MHO catalogue.

\begin{figure}
\centering
  \includegraphics[width=8cm]{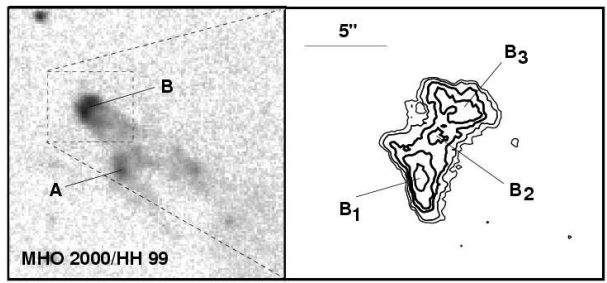}
\caption[]{A simple example of how knots and sub-knots within a single MHO
should be labelled, using letters and, for the sub-knots, numbers.}
 \label{hh99}
\end{figure}

\begin{figure*}
\centering
  \includegraphics[width=14cm]{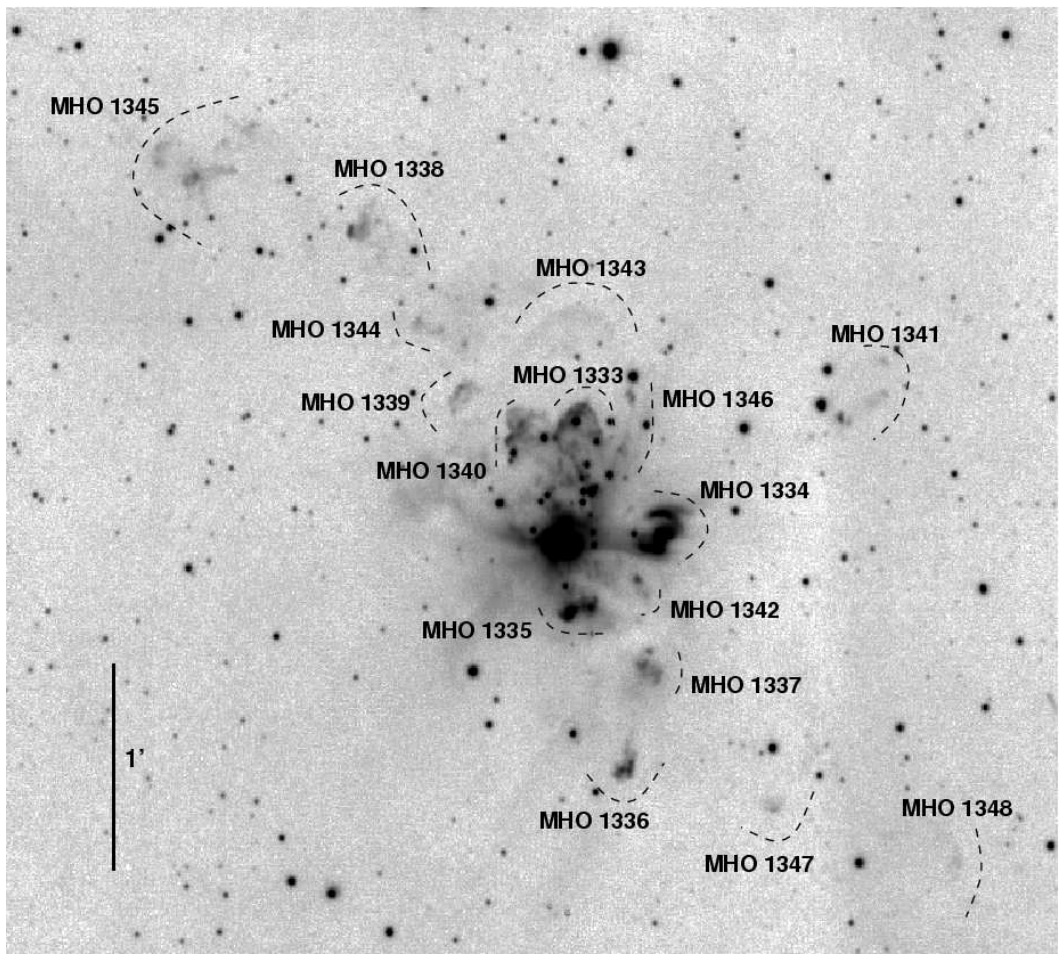}
\caption[]{\htwo\ (+ continuum) image of AFGL 961 in the Rosette
nebula star forming region in Monoceros.  Catalogued MHOs are
labelled; unpublished data obtained with WFCAM at UKIRT (see Davis et
al. 2009 for details of this instrument, the WFCAM data archive and
data processing techniques used to create this image).  }
 \label{rose}
\end{figure*}

Examples of MHOs are shown in Figures~\ref{ori}, \ref{hh99} and
\ref{rose}.  In most cases we have labelled ``groups of knots" rather than
individual features or whole outflows.  Assigning an MHO number to
every resolved feature would of course lead to a vast catalogue that
was impossible to maintain.  On the other hand, associating
widely-spaced knots with a single outflow is often difficult, given
the variability of these line emission features and the large sizes of
some outflows. MHO 187-189 (shown in Figure~\ref{ori}) is a good
example, where three complex groups of features that may well form
part of the same outflow are none-the-less catalogued separately,
although individual knots within each region are not. As with HH
objects, if necessary, individual knots should be identified with
letters; sub-knots should then be labelled with letters and
numbers. HH~99 (MHO~2000) is shown as an example in Figure~\ref{hh99}
(see also the labelling of the knots and sub-features in the detailed,
proper-motion study of HH~47/46 in Eisl\"offel \& Mundt 1994).

In some regions multiple knots and bow shocks radiate in many
directions from a tight cluster of young stars.  Since the
relationship between these objects is often unclear -- each bow shock
may for example be driven by a different outflow that is powered by a
different protostar in the central region -- we also label these
features separately.  An example of such a region, the spectacular
AFGL 961 massive star forming cluster in the Rosette nebula
(described in detail by Aspin 1998 and Li et al. 2008), is shown in
Figure~\ref{rose}.

Finally, for completeness we have also given a catalogue number to
many well-known HH objects (e.g. HH 1/2 = MHO 120/125, HH 212 = MHO
499), though only if these are detected in the near-IR lines of
molecular hydrogen. Whenever possible, we group features together in
a manner consistent with the HH object catalogue.

\section{The MHO Catalogue}

\subsection{Grouping MHOs by region}

There are already almost 1000 objects in the MHO catalogue.  In an
attempt to bring some semblance of order to the list, we have grouped
objects by ``region''.

Strictly speaking, there are no official names for, or boundaries to,
the star-forming giant molecular clouds in our Galaxy.  We have
therefore attempted to define large regions based on the well-defined
boundaries of the 88 constellations (as outlined by the
IAU\footnote{http://www.iau.org/public\_press/themes/constellations/}).
MHOs are almost exclusively confined to molecular clouds in and around
the Gould Belt and the Galactic Plane (the vast majority of molecular
outflows are driven by embedded protostars [Davis et al. 2008, 2009];
relatively few T Tauri stars drive jets that have been detected in
molecular hydrogen line emission, and of course H$_2$ emission, by its
very nature, requires the presence of dense molecular gas).  We have
therefore, in some areas, modified these boundaries slightly to
include large groups of clouds.  We use the large-scale CO J=1-0
survey of the Milky Way, obtained with 1.2 m telescopes in Cambridge,
Massachusetts and Cerro Tololo, Chile (Dame, Hartmann \& Thaddeus
2001) to identify these clouds.  Even so, the boundaries will still
pass through some smaller, less massive clouds and so the boundaries
should only be considered accurate to within a few arcminutes.

The regions defined in this way are listed in the first column in
Table~\ref{regions}.  Note that, in the heavily-populated area of
Orion, we have split the region up into two sub-regions, Orion A and
Orion B, as is the popular convention.

\begin{table*}
\centering
     \caption[]{Regions used to group MHOs}
     \label{regions}

\begin{tabular}{lcccc}
\hline
  \noalign{\smallskip}
  Region$^a$ & Map$^a$ & Approx. RA Range$^b$ & Approx. Dec Range$^b$ & MHO Numbers$^c$\\
  \noalign{\smallskip}
\hline
\noalign{\smallskip}

Perseus   	& M2 & 	03h 00m$\rightarrow$ 04h 00m &	+25\dg$\rightarrow$ +35\dg & 500-699  	\\
Auriga          & M2 &	03h 30m$\rightarrow$ 06h 30m &	+30\dg$\rightarrow$ +56\dg & 1000-1099 	\\
Taurus          & M2 &	03h 00m$\rightarrow$ 05h 50m &	+10\dg$\rightarrow$ +30\dg & 700-799 	\\
Camelopardalis 	& M1 &	04h 00m$\rightarrow$ 08h 00m &	+56\dg$\rightarrow$ +90\dg & 1100-1199 	\\
Orion A 	& M3 &	04h 45m$\rightarrow$ 06h 00m &	-15\dg$\rightarrow$ -04\dg & 1-299 	\\
Orion B         & M3 &	04h 45m$\rightarrow$ 06h 00m &	-04\dg$\rightarrow$ +16\dg & 300-499 	\\
Gemini 	        & M3 &	05h 50m$\rightarrow$ 08h 00m &	+14\dg$\rightarrow$ +34\dg & 1200-1299 	\\
Monoceros 	& M3 &	06h 00m$\rightarrow$ 08h 30m &	-13\dg$\rightarrow$ +14\dg & 1300-1399 	\\
Puppis 	        & M4 &	06h 30m$\rightarrow$ 09h 00m &	-38\dg$\rightarrow$ -13\dg & 1400-1499 	\\
Vela 	        & M4 &	07h 30m$\rightarrow$ 11h 00m &	-55\dg$\rightarrow$ -38\dg & 1500-1599 	\\
Carina 	        & M5 &	08h 00m$\rightarrow$ 12h 00m &	-75\dg$\rightarrow$ -55\dg & 1600-1699 	\\
Chameleon       & M5 &	08h 00m$\rightarrow$ 14h 00m &	-85\dg$\rightarrow$ -70\dg & 3000-3099 	\\
Centaurus 	& M5 &	12h 00m$\rightarrow$ 15h 00m &	-70\dg$\rightarrow$ -30\dg & 1700-1799 	\\
Circinus/Lupus 	& M6 &	15h 00m$\rightarrow$ 16h 00m &	-70\dg$\rightarrow$ -30\dg & 1800-1899 	\\
Scorpius 	& M6 &	16h 00m$\rightarrow$ 18h 00m &	-60\dg$\rightarrow$ -30\dg & 1900-1999 	\\
Corona Australis& M6 &	18h 00m$\rightarrow$ 19h 30m &	-45\dg$\rightarrow$ -35\dg & 2000-2099 	\\
Ophiuchus       & M6 &	16h 00m$\rightarrow$ 18h 00m &	-30\dg$\rightarrow$ +05\dg & 2100-2199 	\\
Serpens 	& M7 & 	17h 30m$\rightarrow$ 18h 40m &	-15\dg$\rightarrow$ +05\dg & 2200-2299 	\\
Sagittarius     & M7 & 	18h 00m$\rightarrow$ 20h 30m &	-35\dg$\rightarrow$ -12\dg & 2300-2399 	\\
Aquila          & M7 &	18h 40m$\rightarrow$ 20h 30m &	-12\dg$\rightarrow$ +15\dg & 2400-2499 	\\
Lyra 	        & M8 &	18h 20m$\rightarrow$ 19h 00m &	+05\dg$\rightarrow$ +45\dg & 2500-2599 	\\
Vulpecula 	& M8 &	19h 00m$\rightarrow$ 21h 30m &	+15\dg$\rightarrow$ +30\dg & 2600-2699 	\\
Cygnus 	        & M9 &	19h 00m$\rightarrow$ 22h 00m &	+30\dg$\rightarrow$ +55\dg & 800-999 	\\
Cepheus      	& M9 &	19h 00m$\rightarrow$ 23h 30m &	+55\dg$\rightarrow$ +90\dg & 2700-2799 	\\
Andromeda 	& M9 &	22h 00m$\rightarrow$ 00h 00m &	+30\dg$\rightarrow$ +55\dg & 2800-2899 	\\
Cassiopeia      & M1 &	23h 00m$\rightarrow$ 04h 00m &	+50\dg$\rightarrow$ +90\dg & 2900-2999 	\\

  \noalign{\smallskip}
  \hline
  \end{tabular}
     
\begin{list}{}{}
\item[$^{\mathrm{a}}$]The name of each region, and the map used to define each region  \\
\item[$^{\mathrm{b}}$]Approximate RA and Dec range associated with each region (a more precise range
           is drawn on each map in Figures~\ref{maps}. \\
\item[$^{\mathrm{c}}$]The range of MHO numbers used for objects within each region (note that
                   objects have not yet been assigned to all numbers in each range). \\
\end{list}
   \end{table*}

\begin{figure*}
\centering
  \includegraphics[width=18cm]{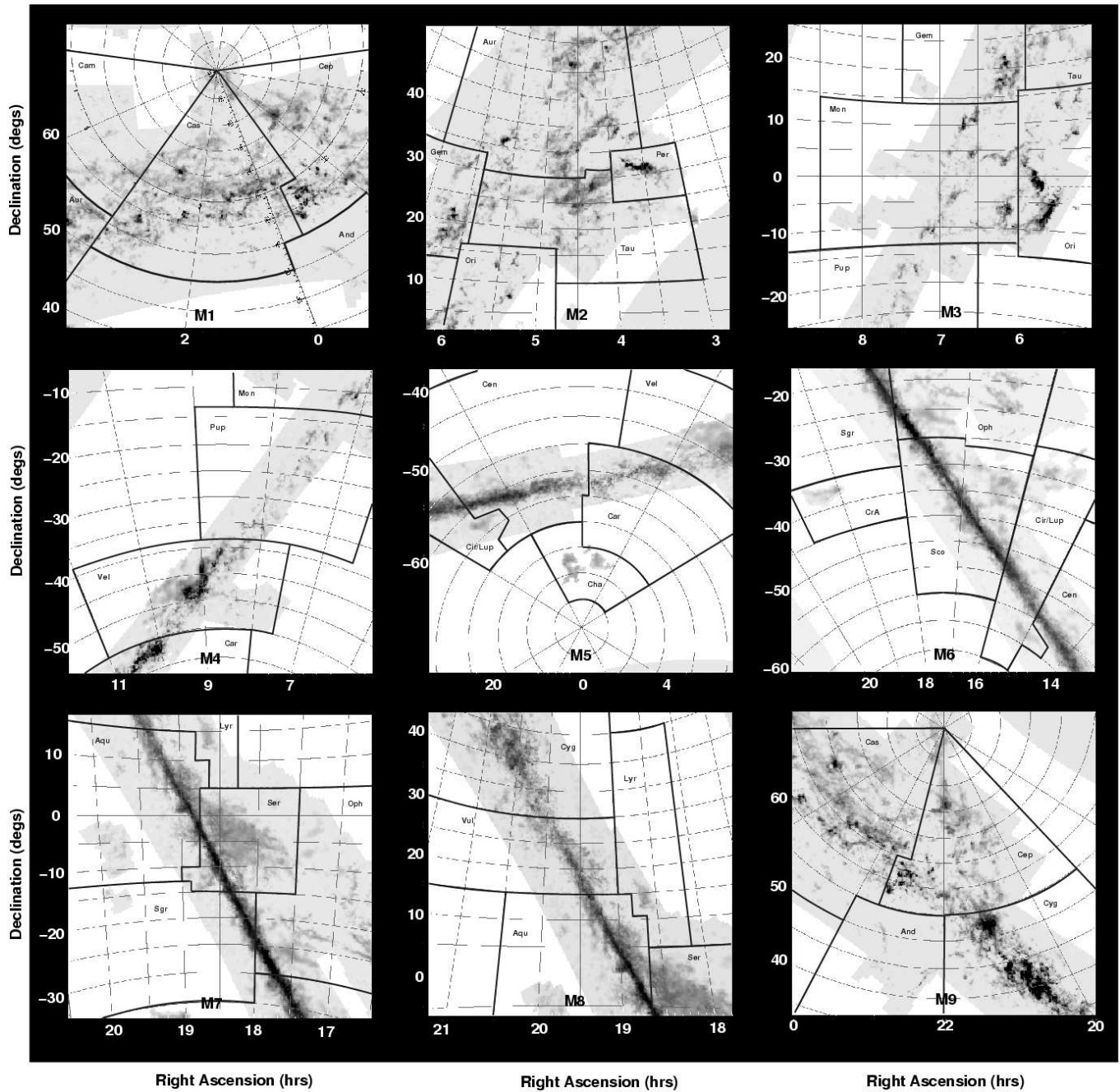}
\caption[]{Large-scale maps in CO J=1-0 emission with the boundaries of the
 regions used to group MHOs marked with thick lines. }
 \label{maps}
\end{figure*}

The boundaries of each region are also marked on low-resolution CO
J=1-0 maps in Figure~\ref{maps}.  M1-M9 in these
figures and in Table~\ref{regions} refer to maps 1 to 9. Note that not
all 88 constellations are listed in Table~\ref{regions}, since those
at high galactic latitudes do not contain star forming regions and/or
known outflows with MHOs.  Indeed, five regions; Camelopardalis,
Centaurus, Circinus/Lupus, Lyra and Andromeda, as yet contain no MHOs.
We include these regions in the catalogue to facilitate the addition
of future discoveries.

The MHO number range listed in the final column in Table~\ref{regions}
defines the range of MHO numbers currently being used in each region.
To date, not all numbers have been assigned to an MHO (in any of the
regions).

The latest version, at the time of writing, of the MHO catalogue
is published here in Appendix A.

\subsection{The on-line database of MHOs}

The entire catalogue is also available on-line at

http://www.jach.hawaii.edu/UKIRT/MHCat.  

\begin{figure*}
\centering
  \includegraphics[width=17cm]{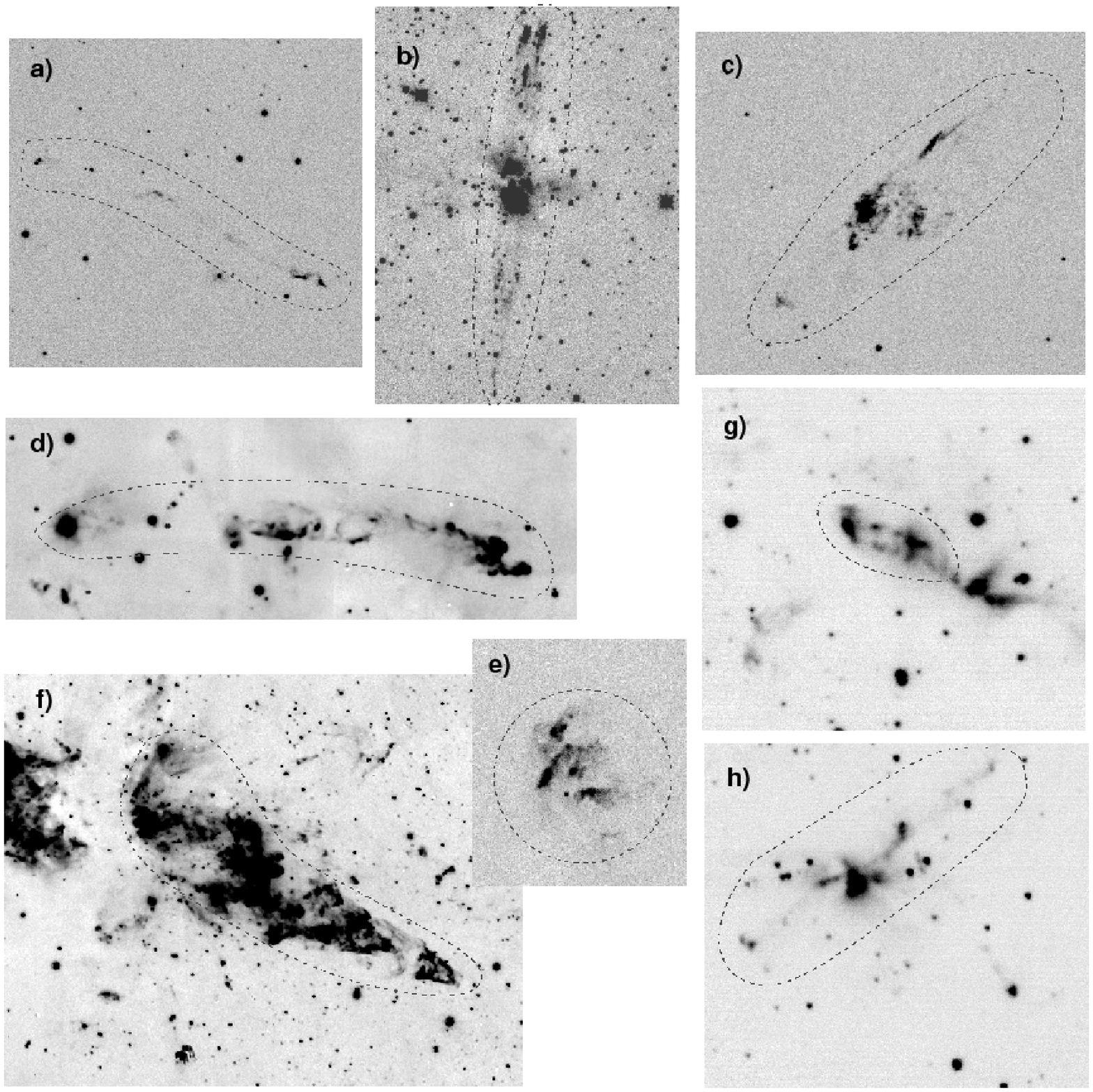}
\caption[]{Examples of the small images available at the MHO web site;
in each case the MHO is marked with a red dashed ellipse or circle:
a) MHO 1300, a curving, collimated jet $\sim$5$'$ SW of the main Mon R2 star 
   forming region in Monoceros (from Hodapp 2007);
b) MHO 1510, a bipolar outflow associated with the bright, nebulous source IRS 20 in 
   Vela (Giannini et al. 2007);
c) MHO 558 (HH 773), a bright, knotty feature in a bipolar molecular outflow
   in the B1 ridge in Perseus (Walawender, Reipurth \& Bally 2009);
d) MHO 18, a spectacular, knotty outflow in the OMC 2/3 region in Orion A 
   (Yu et al. 1997);  
e) MHO 3000, arcs of emission associated with HH 54 in Chameleon (Zealey, Sutters 
   \& Randall 1993); 
f) MHO 899, the luminous south-western molecular flow lobe associated with DR 21 
   in Cygnus (Davis \& Smith 1996);
g) and h) the collimated outflows MHO 2604 and MHO 2201, associated with the
   high-mass star forming regions IRAS 19410+2336 in Vulpecula and IRAS 18151-1208 
   in Serpens, respectively (Varricatt et al. 2010).}
 \label{gifs}
\end{figure*}

\noindent This MHO homepage includes the table of regions shown here
in Table~\ref{regions}; in the on-line catalogue, links in the first
column point to separate tables of MHOs for each region.  These tables
list the MHO number, Right Ascension and Declination, citations to the
discovery paper and subsequent near-IR imaging papers, together with
identifications used in the literature, any associated HH objects, and
a brief description of each object.  A small image of the MHO is also
presented with the object clearly marked; example images from the
on-line catalogue are shown here in Figure~\ref{gifs}.

In the on-line catalogue simple ascii tables are also available.
These list only MHO number, Right Ascension and Declination,
associated HH object, and region.  These very basic tables may be
downloaded and used to plot positions of MHOs on images or maps taken
at other wavelengths, or to label H$_2$ emission-line features in new
near-IR images of star forming regions already covered by the catalogue.

\subsection{Searching through the catalogue}

An easy way to navigate through the catalogue and, in particular, to
search for objects by Right Ascension and Declination, was thought to
be desirable. A $Perl$ script has therefore been developed which
allows the user to enter coordinates and a search radius; the script
returns an HTML table containing MHOs found within the search area. As
with the full region tables, coordinates, references to published
observations, a small image and a brief description of each object is
returned.  This tool is particularly useful for finding MHOs in a star
forming cloud or cluster being studied at different wavelengths, or
for establishing whether an object is a new discovery, or has in fact
already been observed.

\subsection{Checking the catalogue}

Duplicating existing entries and errors associated with the
coordinates assigned to each MHO were our two main concerns when
compiling the catalogue.  To combat both problems, the ascii text
files created for each region were imported into the {\em STARLINK
GAIA} graphical display tool (Draper et al. 2008) and plotted
over wide field R-band Digitised Sky Survey (DSS) images or, if
available, astrometrically-calibrated infrared images.  The infrared
images were all obtained from the UKIRT WFCAM
archive\footnote{http://surveys.roe.ac.uk/wsa/index.html}.

\subsection{The Future}

Our aim is to keep the MHO catalogue as up-to-date as possible. Also,
obviously we want to avoid duplication of catalogue numbers (people
using the same numbers for different objects). Therefore, we ask that
those with new observations please check the catalogue for previous
observations, and contact the catalogue organisers (currently Chris
Davis: c.davis@jach.hawaii.edu) before papers are written, and
certainly before figures and tables of MHOs are finalised, so that new
numbers can be assigned.   

\section{Summary}

A catalogue of molecular hydrogen emission-line objects (MHOs) has
been compiled from the literature. The catalogue includes objects
imaged in molecular hydrogen line emission (almost entirely in the
1-0~S(1) line at 2.122 \mic ).  It does not include objects observed
{\em only} at UV or mid-IR wavelengths.

The catalogue lists only shock-excited features associated with
outflows from young stars.  Objects in both low and high-mass star
forming regions are included. Similar objects associated with
proto-planetary nebulae or extra-galactic sources are not
included.

The catalogue currently contains almost 1000 objects.  Some are
well-known Herbig-Haro objects which we have included for
completeness.  The catalogue is available on-line at
http://www.jach.hawaii.edu/UKIRT/MHCat/ . With the help of the star
formation community, we aim to maintain this catalogue for many
years to come, adding new objects as they are discovered.  We also
hope that in the future, the MHO acronym will be used
universally when labelling these enigmatic objects.

\begin{acknowledgements}

We thank the ``Clearing House" of the Commission 5 Working Group
on Designations, particularly the chair, Marion Schmitz, for their
guidance, and the star formation community, especially Bo Reipurth,
for their valuable input. This project would not have been possible
without support from the Joint Astronomy Centre.

\end{acknowledgements}




\begin{appendix}

\section{Tables of MHOs}

In this section we present tables of MHOs separated by region.  The
regions used to group the MHOs together are defined in
Table~\ref{regions} and in Figure~\ref{maps}.
The very latest versions of these tables are also available on-line at:
http://www.jach.hawaii.edu/UKIRT/MHCat/.

\end{appendix}
\end{document}